\documentstyle[prl, aps, amssymb]{revtex}
\begin{document}
\draft
\wideabs{
\title{Systematic $^{63}$Cu NQR and $^{89}$Y NMR Study of 
Spin Dynamics in Y$_{1-z}$Ca$_{z}$Ba$_{2}$Cu$_{3}$O$_{y}$ Across 
Superconductor-Insulator Boundary}
\author{P.M. Singer and T. Imai}
\address{Department of Physics and Center for Materials Science and 
Engineering, M.I.T., Cambridge, MA 02139}
\date{\today}
\maketitle
\begin{abstract}
We demonstrate that spin dynamics in underdoped
Y$_{1-z}$Ca$_{z}$Ba$_{2}$Cu$_{3}$O$_{y}$ for $y \simeq$ 6.0 exhibit 
qualitatively the same behavior to underdoped La$_{2-x}$Sr$_{x}$CuO$_{4}$ 
for an equal amount of hole concentration $p=z/2=x \leq 0.11$. However, 
a {\it spin--gap} appears as more holes are doped
into the CuO$_{2}$ plane by increasing the oxygen concentration to 
$y \simeq 6.5$
for a fixed value of Ca concentration $z$. Our results also suggest 
that Ca doping causes disorder effects that enhance the 
low frequency spin fluctuations.
\end{abstract}
}

The mechanism of high temperature superconductivity remains a major 
mystery in condensed matter physics. The fundamental 
difficulty stems from the complexity of the electronic phase diagram, 
particularly in the underdoped region. Earlier $^{63}$Cu NMR 
(Nuclear Magnetic Resonance) and 
NQR (Nuclear Quadrupole Resonance) 
measurements of the nuclear spin-lattice relaxation rate $^{63}1/T_{1}$ led to the discovery of 
the {\it pseudo--gap} phenomenon in the spin excitation spectrum of 
bi-layer (Y,La)Ba$_{2}$Cu$_{3}$O$_{y}$ 
\cite{Horvatic,Warren,Yasuoka}. In the {\it spin pseudo--gap}, or {\it 
spin--gap} regime, low energy spin excitations are 
suppressed below the spin--gap temperature $T^{*}$ ($>T_{c}$) which
results in a decrease in $^{63}1/T_{1}T$
($^{63}1/T_{1}$ divided by temperature $T$) below $T^{*}$.
Subsequent NMR 
and optical charge transport measurements showed that a 
pseudo--gap appears both in the spin and charge excitation spectrum 
of a wide variety of high $T_{c}$ cuprates
\cite{Schlesinger,Homes,Itoh,Julien,Ishida}, with the most notable exception being the prototype 
high $T_{c}$ cuprate La$_{2-x}$Sr$_{x}$CuO$_{4}$. 
Moreover, the temperature scale $T^{*}$ of the spin--gap and charge--gap increases with decreasing 
hole concentration towards the superconductor--insulator boundary 
\cite{Yasuoka,Homes,Itoh,Ishida}.  

In various theoretical model analysis, the pseudo--gap is often considered 
the key in understanding the mechanism of superconductivity.  Unfortunately, 
driving CuO$_{2}$ planes into the insulating regime in a controlled fashion 
is technically difficult in many high $T_{c}$ cuprates.  As such, the fate 
of the pseudo-gap in the 
heavily underdoped insulating regime has been highly controversial.  
Attempts have been made to infer information on the
spin-gap based on uniform spin susceptibility $\chi'({\bf q}={\bf 0})$ 
\cite{Hwang,Oda}.
However, it is important to realize 
that growth of short range spin order alone causes a reduction of $\chi'({\bf q}={\bf 0})$ 
{\it without} having any gaps. 
For example, the undoped CuO$_{2}$ plane shows a roughly 
linear decrease of $\chi'({\bf q}={\bf 0})$ with decreasing 
temperature which is entirely consistent with the 2-d Heisenberg 
model. By continuity, it is natural to associate the 
decrease  of $\chi'({\bf q}={\bf 0})$ in the heavily underdoped regime 
to be mostly due to growth of short range order \cite{Johnston} and not to a spin-gap.


It has also become increasingly popular to infer $T^{*}$ 
for the charge sector based on scaling analysis of the Hall effect 
\cite{Hwang} or resistivity data \cite{Cooper}. Some authors claim that there is a universal phase 
diagram of $T^{*}$ with $p$ and $T$ being the only two parameters, even 
in La$_{2-x}$Sr$_{x}$CuO$_{4}$. However, earlier $^{63}$Cu NQR 
\cite{Ohsugi,Imai} and neutron scattering \cite{Yamada-PRL} experiments
revealed no hint of a spin--gap above $T_{c}$ in La$_{2-x}$Sr$_{x}$CuO$_{4}$.   
Instead, La$_{2-x}$Sr$_{x}$CuO$_{4}$ exhibits an instability at low temperatures 
towards the formation of the quasi--static {\it stripe} with incommensurate 
spin and charge density waves \cite{Tranquada,Lee,Wakimoto}.
Careful NMR(NQR) experiments of spin--gap effects with controlled 
doping near the superconductor-insulator boundary are necessary and would 
allow for comparison with La$_{2-x}$Sr$_{x}$CuO$_{4}$. 

In this Letter, we report a systematic microscopic investigation of 
Y$_{1-z}$Ca$_{z}$Ba$_{2}$Cu$_{3}$O$_{y}$ utilizing $^{63}$Cu NQR 
and $^{89}$Y NMR.
The advantage of the Y$_{1-z}$Ca$_{z}$Ba$_{2}$Cu$_{3}$O$_{y}$ system 
is that one can control the hole concentration near the superconductor-insulator
boundary by fixing $y \simeq 6.0$ and varying $z$. In this case, the hole 
concentration is given by $p=z/2$, because the chain
Cu sites (Cu(1)) with two-fold oxygen coordination remain insulating with a
3d$^{10}$ configuration \cite{Vega}. In Fig. 1(a) we show the {\it absence} of a
spin-gap signature in
heavily underdoped Y$_{1-z}$Ca$_{z}$Ba$_{2}$Cu$_{3}$O$_{6.0}$
based on measurements of $^{63}1/T_{1}T$ at the planar Cu site (Cu(2)).
Instead, we show that the low energy spin excitations exhibit similar behavior to underdoped 
La$_{2-x}$Sr$_{x}$CuO$_{4}$, with equivalent
$p=z/2=x$, which monotonically grow with decreasing $p$ and $T$.
With further hole doping Y$_{1-z}$Ca$_{z}$Ba$_{2}$Cu$_{3}$O$_{6.0}$ by oxygen loading
to $y \simeq 6.5$ for fixed $z$,
however, we $do$ observe the spin-gap signature (Fig. 1(a)), even though it 
appears somewhat suppressed compared to YBa$_{2}$Cu$_{3}$O$_{6.5}$
without Ca substitution (Fig. 1(b)).
This is the first time in the high $T_{c}$ cuprates where the appearance 
of a spin--gap signature is experimentally tracked through the 
insulator-superconductor boundary by increasing the hole doping. We 
recall that in contrast with the present case, further hole doping 
La$_{2-x}$Sr$_{x}$CuO$_{4}$ (or La$_{2}$CuO$_{4+\delta}$) does 
$not$ result in a spin--gap signature,
therefore our finding challenges the popular argument that 
charge disorder caused by the alloying effects of Sr$^{+2}$ 
substitution (i.e. "dirt effects") {\it alone}
suppresses the spin-gap and drives La$_{2-x}$Sr$_{x}$CuO$_{4}$ towards 
the stripe instability.

We synthesized our polycrystalline samples following \cite{Casalta}. The 
oxygen concentration was controlled and determined following 
\cite{Ueda} with precision $\Delta y \sim \pm$0.05. 
The $^{63}$Cu NQR spectrum 
of all our Y$_{1-z}$Ca$_{z}$Ba$_{2}$Cu$_{3}$O$_{y}$ samples
are very similar to those 
reported earlier by Vega {\it et al.} \cite{Vega}, and the superconducting transition temperature 
$T_{c}$, as determined by SQUID measurements, shows close agreement with \cite{Casalta}.
The temperature dependence of $^{63}1/T_{1}$ was measured with NQR near 
the peak frequency of the $^{63}$Cu(2) site at $\omega_{n}/2\pi \simeq 25.5$ MHz \cite{Vega} by 
applying an inversion pulse prior to the spin echo sequence. A
typical $\pi/2$-pulse width of 3 $\mu$s was used. NMR measurements
at 9 Tesla in uniaxially aligned powder gave 
identical results to NQR within uncertainties.  
$^{63}1/T_{1}T$ is given by
\begin{equation}
        ^{63}  \frac{1}{T_{1}T} = \frac{ 2 k_{B}}{g^{2} \mu_{B}^{2} \hbar^{2}} 
	\sum_{{\bf q}} |^{63}A({\bf q}) |^{2} \frac{\chi''({\bf q},\omega_{n})}{\omega_{n}}
\label{T1}
\end{equation}
where $^{63}A({\bf q})$ is the wave-vector dependent, geometrical form factor
of the electron-nucleus hyperfine 
coupling \cite{Moriya,MMP}.
As shown in Fig. \ref{T1T}(a), $^{63}1/T_{1}T$ in underdoped 
Y$_{1-z}$Ca$_{z}$Ba$_{2}$Cu$_{3}$O$_{6.0}$ ($z\leq 0.22$) with nominal 
hole concentration $p\leq 0.11$ does {\it not} exhibit a spin--gap.
Instead, $^{63}1/T_{1}T$
grows with decreasing temperature, exhibiting similar 
values as La$_{2-x}$Sr$_{x}$CuO$_{4}$
for the equivalent hole concentration $p=z/2=x$ shown in Fig. 
\ref{89Y}(a).
The fact that $^{63}1/T_{1}T$ 
grows with decreasing temperature indicates that 
low energy spin excitations
continue to increase with decreasing temperature. 
Moreover, the enhancement of low energy spin 
excitations below 300 K is followed by the decrease of the
$^{63}$Cu NQR signal intensity below $T_{wipeout} (\gtrsim 
200$ K), i.e. {\it wipeout} effects \cite{Hunt,Singer,Hunt2}. 
Wipeout effects can be caused by various 
mechanisms \cite{Hunt} including the presence of nearly localized free spins 
induced by hole localization (in 
analogy with Cu NMR wipeout in Cu metal imbedded with 
dilute Fe or Mn spins), as well as the onset of the
glassy slowing down of stripes. We note that as a consequence of
wipeout effects,
the value of $^{63}1/T_{1}T$ measured below $T_{wipeout}$ does {\it not} 
represent that of the entire CuO$_{2}$ plane.

The temperature dependence of $\chi'({\bf q}={\bf 0})$ was deduced 
from the spin contribution 
$^{89}K_{spin}=D\chi'({\bf q}={\bf 0})$ to the $^{89}$Y NMR Knight shift, 
\begin{equation}
          ^{89}K = ^{89}K_{orb}+^{89}K_{spin}
\label{T1}
\end{equation}
as shown in Fig. \ref{89Y}(b), where the powder averaged orbital contribution is
$^{89}K_{orb}=+150 \pm 5$ ppm \cite{Alloul} and $D$ ($< 0$) is the hyperfine coupling 
constant. Our $^{89}K$ data, taken in a 
magnetic field of 9 Tesla, for 
Y$_{0.78}$Ca$_{0.22}$Ba$_{2}$Cu$_{3}$O$_{y}$ are consistent with 
earlier results reported above $\sim$110 K for
Y$_{0.8}$Ca$_{0.2}$Ba$_{2}$Cu$_{3}$O$_{y}$ 
by Williams {\it et al.} \cite{Williams} also shown in Fig. \ref{89Y}(b). 
Our new measurement conducted down to $T_{c}=75 $ K 
in Y$_{0.78}$Ca$_{0.22}$Ba$_{2}$Cu$_{3}$O$_{6.9}$ shows clear 
signature of saturation {\it below} 100 K, 
similar to overdoped YBa$_{2}$Cu$_{3}$O$_{7}$ without Ca substitution 
\cite{Auler}. The saturation of $\chi'({\bf q}={\bf 0})$ below 100 K is followed by a broad maximum at 
$T_{max} = 90 \pm 10$ K, which according to the 2-d Heisenberg model 
\cite{Singh}, implies an effective energy scale $J(p)=T_{max}/0.93 = 97\pm 11$ K. 
We also deduce $J(p)$ in underdoped
Y$_{0.78}$Ca$_{0.22}$Ba$_{2}$Cu$_{3}$O$_{6.0}$ and 
Y$_{0.78}$Ca$_{0.22}$Ba$_{2}$Cu$_{3}$O$_{6.5}$ by matching $\chi'({\bf 
q}={\bf 0})$ to the low 
temperature ($T \ll J(p)$) portion of the 2-d Heisenberg model 
\cite{Johnston}, as shown in Fig. \ref{89Y}(b).
Our results of 
$J(p)$ summarized in Fig. \ref{Phase} are consistent with those reported in 
La$_{2-x}$Sr$_{x}$CuO$_{4}$ \cite{Johnston}.

The $^{89}$Y NMR data also shows evidence for the glassy slowing of disordered magnetism.
Below the onset of glassy slowing at $T_{wipeout} 
(\gtrsim 200$ K), we find a change  in
curvature of $^{89}1/T_{1}T$ and an increase in $^{89}\Delta 
f$, as shown in Fig. \ref{89Y}(c) and (d) respectively \cite{Carretta}. The change in 
curvature of $^{89}1/T_{1}T$ is first followed by a minimum at 
$T_{89}^{min}$ \cite{K}, and then a maximum at
$T_{89}^{max}$ where the glassy slowing has reached 
the NMR time--scale.
At a similar temperature $T_{\mu SR}$, $\mu$SR measurements
observe local hyperfine fields \cite{Niedermayer} that are frozen on 
the $\mu$SR time--scale. The enhanced values of $^{89}\Delta f$ at
1.7 K also indicate that the frozen 
hyperfine fields at the $^{89}$Y nuclear site have a substantial spatial distribution $\sim 70$ Oe.

The sequence of anomalies starting at $T_{wipeout}$ followed by
$T_{89}^{min}$, $T_{89}^{max}$ and $T_{\mu SR}$ (all shown in Fig. 
\ref{Phase}) are analogous to La$_{2-x}$Sr$_{x}$CuO$_{4}$
where the on-set of glassy slowing down of the stripe phase 
at $T_{wipeout}$ \cite{Hunt,Singer,Hunt2} is 
followed by a minimum then a maximum
in $^{139}1/T_{1}T$ \cite{Chou} and $\mu$SR observation of 
frozen hyperfine fields \cite{Niedermayer}.
These set of results establish the 
following three points: First, the paramagnetic Cu spin fluctuations in 
underdoped CuO$_{2}$ planes 
exhibit nearly universal $p$ and $T$ dependences in
Y$_{1-z}$Ca$_{z}$Ba$_{2}$Cu$_{3}$O$_{6.0}$ and
La$_{2-x}$Sr$_{x}$CuO$_{4}$ for equivalent $p=z/2=x$, {\it 
without} a spin--gap signature.  
In the same temperature range, the $^{89}$Y NMR Knight shift decreases 
monotonically and is most likely due to the 
growth of short range spin order.
Second, the gradual slowing of Cu 
spin fluctuations, as observed by the increase in 
$^{63}1/T_{1}T$, is followed by glassy freezing of the Cu moments 
starting at $T_{wipeout} \sim$ 200 K in 
Y$_{1-z}$Ca$_{z}$Ba$_{2}$Cu$_{3}$O$_{6.0}$ while similar behavior 
is observed only below $\sim$100 K in 
La$_{2-x}$Sr$_{x}$CuO$_{4}$ within a similar doping range.
The factor 2 higher temperature scale is consistent with the finding 
based on $\mu$SR that the
spin freezing temperature $T_{\mu SR}$ in 
Y$_{1-z}$Ca$_{z}$Ba$_{2}$Cu$_{3}$O$_{6.0}$ is also a factor $\sim$2 
higher than in La$_{2-x}$Sr$_{x}$CuO$_{4}$ \cite{Niedermayer}. Recalling that the 
N\'{e}el temperature of $T_{N}$=420 K in undoped YBa$_{2}$Cu$_{3}$O$_{6.0}$ is 
higher than $T_{N}$=320 K in La$_{2}$CuO$_{4}$ because of the bi-layer 
coupling, the higher temperature scale of glassy spin freezing in Y$_{1-z}$Ca$_{z}$Ba$_{2}$Cu$_{3}$O$_{6.0}$ 
may also be due to the stronger 3-d coupling along the 
$c$-axis. However, we cannot rule 
out the possibility that Ca$^{+2}$ substitution causes stronger disorder in 
Y$_{1-z}$Ca$_{z}$Ba$_{2}$Cu$_{3}$O$_{6.0}$ than 
Sr$^{+2}$ substitution in La$_{2-x}$Sr$_{x}$CuO$_{4}$, as suggested by 
the factor $\sim 2$ broader $^{63}$Cu NQR spectrum \cite{Vega}, which may 
enhance the tendency towards spin freezing. Third, the observed increase of 
$^{89}1/T_{1}T$ implies that the Cu moments are not 
slowing down towards the commensurate 
antiferromagnetic spin structure with divergently large spin-spin correlation 
length. In this context, it is important to recall that the critical 
slowing down towards the N\'{e}el state does not cause a large 
enhancement of $^{89}1/T_{1}T$ in undoped YBa$_{2}$Cu$_{3}$O$_{6.0}$ \cite{Ohno}
since the hyperfine form factor $^{89}A({\bf q})$ is zero for the commensurate wave vectors. 
The strong increase of $^{89}1/T_{1}T$ shows that either 
the spin structure is incommensurate, as expected for the stripe phase, or that the spin-spin 
correlation length is limited to a relatively short length scale due 
to disorder caused by the holes, or possibly both. Stripes are 
dynamic at NMR time scales even at $\sim 350 $ mK as evidenced by 
motional narrowing effects \cite{Hunt2}, therefore the exact spin 
configuration cannot be distinguished using NMR.

We have established that the slowing of the paramagnetic spin 
dynamics in Y$_{1-z}$Ca$_{z}$Ba$_{2}$Cu$_{3}$O$_{6.0}$ is 
qualitatively similar to La$_{2-x}$Sr$_{x}$CuO$_{4}$. Most 
importantly, we do not observe the signature of a spin gap. Instead, we 
find signatures of glassy slowing of spin fluctuations similar to the 
case of La$_{2-x}$Sr$_{x}$CuO$_{4}$. 
We caution that the absence of a spin--gap signature in the form of a decrease 
in $^{63}1/T_{1}T$ does not necessarily {\it prove} that there is no 
global suppression of lower energy parts of the spin fluctuations.  
$^{63}1/T_{1}T$ may grow monotonically with decreasing temperature as 
long as very low frequency ($\sim \omega_{n}$) components of the spin fluctuations 
grow, even if the global spin fluctuation spectrum is gapped below a 
certain temperature $T^{*}$. On the other hand, 
$T_{wipeout}$ sets an upperbound on 
$T^{*}$ in Y$_{1-z}$Ca$_{z}$Ba$_{2}$Cu$_{3}$O$_{6.0}$ because if $T^{*}$ is 
significantly larger than $T_{wipeout}$, we should observe the 
decrease of $^{63}1/T_{1}T$ prior to the influence of glassy slowing of 
the spin dynamics which become visible below $T_{wipeout}$. Our finding 
that the magnitude of $T^{*}$ is at most comparable to $T_{wipeout}$ 
in Y$_{1-z}$Ca$_{z}$Ba$_{2}$Cu$_{3}$O$_{6.0}$ 
is at odds with popularly held speculations, often based on 
theoretical expectations or more indirect experimental information 
such as the Hall effect, resistivity, and $\chi'({\bf q}={\bf 0})$, that $T^{*}$ blows up 
towards $J(p=0)\sim 1500$ K.

A potential common cause of the absence of the spin--gap signature in underdoped
Y$_{1-z}$Ca$_{z}$Ba$_{2}$Cu$_{3}$O$_{6.0}$ and
La$_{2-x}$Sr$_{x}$CuO$_{4}$ is the random charge potential and/or
disorder induced by substitution of Ca$^{+2}$ or Sr$^{+2}$ ions into
Y$^{+3}$ or La$^{+3}$ sites respectively. It is worth recalling that
the absence of a spin--gap signature in La$_{2-x}$Sr$_{x}$CuO$_{4}$ has
often been attributed to ``dirt effects'' caused by Sr$^{+2}$.
However, our results in Fig. \ref{T1T}(a) also indicate
that disorder {\it alone} does not entirely suppress the spin--gap. Due to
the solubility limit of Ca$^{2+}$ into
Y$_{1-z}$Ca$_{z}$Ba$_{2}$Cu$_{3}$O$_{6.0}$ with a maximum $T_{c} \sim 
30$ K \cite{Casalta},
we doped more holes into Y$_{0.78}$Ca$_{0.22}$Ba$_{2}$Cu$_{3}$O$_{6.0}$
by adding oxygen into the chain layers for the {\it same} sample
to obtain Y$_{0.78}$Ca$_{0.22}$Ba$_{2}$Cu$_{3}$O$_{6.50}$ with $T_{c}=59$ K. We found that
$^{63}1/T_{1}T$ in Y$_{0.78}$Ca$_{0.22}$Ba$_{2}$Cu$_{3}$O$_{6.50}$
decreases below $T^{*}\sim 130$ K, similar to the
spin--gap signature in
YBa$_{2}$Cu$_{3}$O$_{6.50}$ \cite{Yasuoka,ImaiData}. 
The data therefore suggests that the spin--gap {\it does} develop when more holes are added 
into the CuO$_{2}$ plane in 
Y$_{0.78}$Ca$_{0.22}$Ba$_{2}$Cu$_{3}$O$_{y}$, even if Ca doping tends to suppress the spin--gap 
signature.

In order to test the effects of
Ca substitution in a 
more systematic fashion, we compare $^{63}1/T_{1}T$ 
for Y$_{1-z}$Ca$_{z}$Ba$_{2}$Cu$_{3}$O$_{6.5\sim 6.55}$ with $z=0$, 
0.08, and 0.22 (Fig. \ref{T1T}(b)).
$^{63}1/T_{1}T$ is systematically enhanced with increasing $z$, especially at lower 
temperatures. Our data suggest that Ca$^{+2}$ doping not 
only introduces holes but also gives rise to disorder effects which tend to fill in the low frequency parts of 
spin fluctuation spectrum, without affecting the 
magnitude of $T^{*}$ significantly.
The Ca
substitution effects for $y \simeq 6.5$ is in remarkable 
contrast with the lack of change in $^{63}1/T_{1}T$
observed for $y\simeq 6.9$ (Fig. 1). Our results for $y\simeq 6.9$ are 
consistent with earlier reports \cite{Williams2}.
It is interesting to note the qualitative similarity with the Zn substitution effects 
in YBa$_{2}$Cu$_{3}$O$_{y}$\cite{Mahajan}.
$^{89}$Y NMR by Mahajan {\it et al.} \cite{Mahajan} showed that Zn 
substitution causes
$^{89}$Y line splitting in $T_{c} \simeq$60 K phase samples but
causes only $^{89}$Y NMR line broadening in the $T_{c} \simeq 90$ K phase 
with $y\simeq 6.9$. 
These results suggest that both random spinless 
impurities in the CuO$_{2}$ plane (Zn$^{2+}$) and random Coulomb potentials 
outside the CuO$_{2}$ plane (Ca$^{2+}$) are more effectively shielded 
by a larger number of holes in the overdoped region. We mention 
that a more detailed analysis of the $^{63}$Cu(2) spin-lattice 
recovery, similar to that used for Zn doped
YBa$_{2}$Cu$_{4}$O$_{8}$ by Itoh {\it et al.} \cite{ItohZn}, is unfortunately not 
possible in Y$_{1-z}$Ca$_{z}$Ba$_{2}$Cu$_{3}$O$_{y}$
due to the small overlap ($\sim 1 \%$) of the Cu(1) signal with 
very long $^{63}T_{1}$ \cite{Vega}.

To conclude, using both $^{63}$Cu NQR and $^{89}$Y NMR we
demonstrate the remarkable 
similarity in the paramagnetic spin dynamics between 
Y$_{1-z}$Ca$_{z}$Ba$_{2}$Cu$_{3}$O$_{6.0}$ and La$_{2-x}$Sr$_{x}$CuO$_{4}$ 
for equivalent nominal hole concentration.
We do not observe any signatures of a spin--gap for $p=z/2=x\leq 0.11$.
Upon further hole doping by oxygen loading with fixed $z$, we demonstrate that a
spin--gap $does$ develop. Combining all our data, we deduce a phase diagram which crosses the
superconductor--insulator boundary and
includes the spin--gap temperature $T^{*}$, the effective energy scale 
$J(p)$ ($\gg T^{*}$), 
and the glassy freezing of the spin dynamics. Our systematic study 
of Ca substitution suggest that charge disorder caused by 
Ca$^{+2}$ ions tends to suppress the spin--gap signature while keeping 
$T^{*}$ ($< T_{wipeout}$) roughly constant.

T.I. thanks M. Greven, C. Nayak, S. Chakravarty, and X.-G. Wen for inspiring this project. 
This work was supported by NSF DMR 99-71264 and NSF DMR 98-08941.

\begin{figure}
\caption{(a) $^{63}1/T_{1}T$ above $T_{c}$ (vertical lines) at the 
$^{63}$Cu(2) site for
Y$_{1-z}$Ca$_{z}$Ba$_{2}$Cu$_{3}$O$_{y}$ (where $[z$ $y$ $T_{c}]$
is indicated in (b)) and for 
La$_{1.885}$Sr$_{0.115}$CuO$_{4}$ (dotted curve). 
($\blacksquare$) and ($\square$) are taken from [20]. All lines are a guide 
for the eye. Dashed line through ($\blacktriangle$) indicates region below 
$T_{wipeout}$ where only partial 
Cu(2) signal intensity exists.}
\label{T1T}
\end{figure}

\begin{figure}
\caption{The same symbol assignment as Fig. 1 is used, and new 
symbols are shown. (a) $^{63}1/T_{1}T$ 
in Y$_{1-z}$Ca$_{z}$Ba$_{2}$Cu$_{3}$O$_{y}$ 
and La$_{2-x}$Sr$_{x}$CuO$_{4}$.
(b) $\chi'({\bf q=0})$ in Y$_{1-z}$Ca$_{z}$Ba$_{2}$Cu$_{3}$O$_{y}$ as measured by the $^{89}$Y NMR Knight shift 
($^{89}K$)
taken above $T_{89}^{min}$ [28] with respect to a YCl$_{3}$ reference.
Data ($\times$) from [29] are a series of
$[0.20$ $y$ $T_{c}]$ samples with 
$T_{c}$ = 47.5 K, 65.8 K, 83.2 K, 86 K, 72.1 K, 60 K, and 47.5 K (in order of 
increasing $|^{89}K_{spin}|$).
Arrow indicates net spin contribution $^{89}K_{spin}$. Solid lines are fits to the
2-d Heisenberg model [30], and all other lines in figure are guides for the eye.
(c) $^{89}1/T_{1}T$ (with same data plotted below 30 K in the inset) and
(d) full width at half maximum $^{89}\Delta f$ of the $^{89}$Y NMR 
line--shape.}
\label{89Y}
\end{figure}

\begin{figure}
\caption{Phase diagram of 
Y$_{1-z}$Ca$_{z}$Ba$_{2}$Cu$_{3}$O$_{y}$ as a function of 
Ca$_{z}$ substitution for fixed $y \simeq 6.0$ to the 
left of dashed vertical line, and as a function of O$_{y}$ concentration
for fixed $z=0.22$ to the right.
The data includes $T_{N}$ ($\blacklozenge$) [21], $T_{\mu SR}$ ($\blacktriangle$) 
[33], $T_{89}^{min}$
($\blacktriangledown$), $T_{89}^{max}$ ($\triangledown$),  
$T_{wipeout}$ (hatched region), $T^{*}$ ($\bullet$),
$T_{c}$ ($\lozenge$) and
$J(p)$ ($\ast$,$\times$) deduced according to $T_{max}$ and
fit to the 2-d Heisenberg model, respectively. 
All data to the right, including $T_{c}$ ($+$) from [29], are 
positioned linearly according to $|^{89}K_{spin}|$ at 300 K. All lines are a guide for the eye and 
($\circ$) is $T^{*}$ for [0.08 6.5 62K].}
\label{Phase}
\end{figure}


\begin{references}
\bibitem{Horvatic}M. Horvati\'{c} {\it et al.}, Phys. Rev. B {\bf 39}, 7332 (1989).
\bibitem{Warren}W.W. Warren {\it et al.}, Phys. Rev. Lett. {\bf 62}, 1193 (1989).
\bibitem{Yasuoka}H. Yasuoka, T. Imai, and T. Shimizu, in {\it Strong 
Correlation and Superconductivity}, Eds. H.
Fukuyama, S. Maekawa, and A. P. Malozemoff, Springer-Verlag (1989).
Also see A. Goto {\it et al.}, Phys. Rev. B {\bf 55}, 12736 (1997) and 
A. Goto {\it et al.}, J. Phys. Soc. Jpn. {\bf 65}, 3043 (1996).
\bibitem{Schlesinger}Z. Schlesinger {\it et al.}, Phys. Rev. Lett. {\bf 65}, 801 (1990).
\bibitem{Homes}C.C. Homes {\it et al.}, Phys. Rev. Lett. {\bf 71}, 1645 (1993).
\bibitem{Itoh}Y. Itoh {\it et al.}, J. Phys. Soc. Jpn. {\bf 63}, 22 (1994) 
and {\it ibid} J. Phys. Soc. Jpn. {\bf 65}, 3751 (1996).
\bibitem{Julien}M.-H. Julien {\it et al.}, Phys. Rev. Lett. {\bf 76}, 4238 (1996).
\bibitem{Ishida}K. Ishida {\it et al.} Phys. Rev. B {\bf 58}, R5960 (1998).
\bibitem{Hwang}H.Y. Hwang {\it et al.} Phys. Rev. Lett. {\bf 72}, 2636 (1994).
\bibitem{Oda}T. Nakano {\it et al.}, J. Phys. Soc. Jpn. {\bf 67}, 2622 (1998).
\bibitem{Johnston}D.C. Johnston, Phys. Rev. Lett. {\bf 62}, 957 (1989).
\bibitem{Cooper}J.R. Cooper {\it et al.}, Physica C {\bf 341}, 855 (2000).
\bibitem{Ohsugi}S. Ohsugi {\it et al.} J. Phys. Soc. Jpn. {\bf 60}, 2351 (1991).
\bibitem{Imai}T. Imai {\it et al.} Phys. Rev. Lett. {\bf 70}, 1002 (1993).
\bibitem{Yamada-PRL}K. Yamada {\it et al.} Phys. Rev. Lett. {\bf 75}, 1626 (1995).
\bibitem{Tranquada}J.M. Tranquada {\it et al.} Nature {\bf 375}, 561 (1995).
\bibitem{Lee}Y.S. Lee {\it et al.} Phys. Rev. B. {\bf 60}, 3643 (1999).
\bibitem{Wakimoto}S. Wakimoto {\it et al.} Phys. Rev. B {\bf 63}, 172501 (2001).
\bibitem{Vega}A.J. Vega {\it et al.}, Phys. Rev. B {\bf 39}, 2322 (1989) 
and {\it ibid} Phys. Rev. B {\bf 40}, 8878 (1989).
\bibitem{ImaiData}T. Imai {\it et al.}, Physica (Amsterdam) {\bf 162C}, 169 (1989). 
\bibitem{Casalta}H. Casalta {\it et al.}, Physica (Amsterdam) {\bf 204C}, 331 (1993).
\bibitem{Ueda}Y. Ueda {\it et al.}, Physica (Amsterdam) {\bf 156C}, 281 (1988).
\bibitem{Moriya}T. Moriya, J. Phys. Soc. Jpn. {\bf 18}, 516 (1963).
\bibitem{MMP}A.J. Millis {\it et al.}, Phys. Rev. B {\bf 42}, 167 (1990). 
\bibitem{Hunt}A.W. Hunt {\it et al.}, Phys. Rev. Lett. {\bf 82}, 4300 
(1999). Also see T. Imai {\it et al.}, J. Phys. Soc. Jpn. {\bf 59}, 3846 (1989).
\bibitem{Singer}P.M. Singer {\it et al.}, Phys. Rev. B {\bf 60}, 15345 (1999).
\bibitem{Hunt2}A.W. Hunt {\it et al.}, Phys. Rev. B {\bf 64}, 134525 (2001).
\bibitem{K}$T_{89}^{min}$ also signals the temperature below 
which $^{89}1/T_{1}T \propto$ $^{89}K_{spin}$ ($\propto$ $^{89}\Delta 
f$) scaling \cite{Alloul} no longer holds.
\bibitem{Williams}G.V.M. Williams {\it et al.}, Phys. Rev. B {\bf 54}, R6909 (1996)
and {\it ibid} Phys. Rev. B {\bf 57}, 8696 (1998).
\bibitem{Singh}R.R.P. Singh {\it et al.}, Phys. Rev. B. {\bf 42}, 996 (1990).
\bibitem{Alloul}H. Alloul {\it et al.}, Phys. Rev. Lett {\bf 70}, 1171 (1993).
\bibitem{Auler}T. Auler {\it et al.}, Phys. Rev. B {\bf 56}, 11294 (1997).
\bibitem{Niedermayer}Ch. Niedermayer {\it et al.}, Phys. Rev. Lett. {\bf 80}, 3843 (1998).
\bibitem{Carretta}Independent $^{89}$Y NMR data in
Y$_{0.85}$Ca$_{0.15}$Ba$_{2}$Cu$_{3}$O$_{6.1}$ by 
A. Campana {\it et al.}, Int. J. Mod. Phys. B {\bf 14}, 2797 (2000) are 
consistent with our data.
\bibitem{Chou}F.C. Chou {\it et al.}, Phys. Rev. Lett. {\bf 71}, 2323 (1993).
\bibitem{Ohno}T. Ohno {\it et al.}, J. Phys. Soc. Jpn. {\bf 60}, 2040 (1991).
\bibitem{Williams2}G.V.M. Williams {\it et al.}, Phys. Rev. B {\bf 63}, 104514 (2001).
\bibitem{Mahajan}A.V. Mahajan {\it et al.}, Euro. Phys. J. B {\bf 13}, 457 (2000).
\bibitem{ItohZn}Y. Itoh {\it et al.}, J. Phys. Soc. Jpn. {\bf 70}, 1881 (2001).
\end{references}
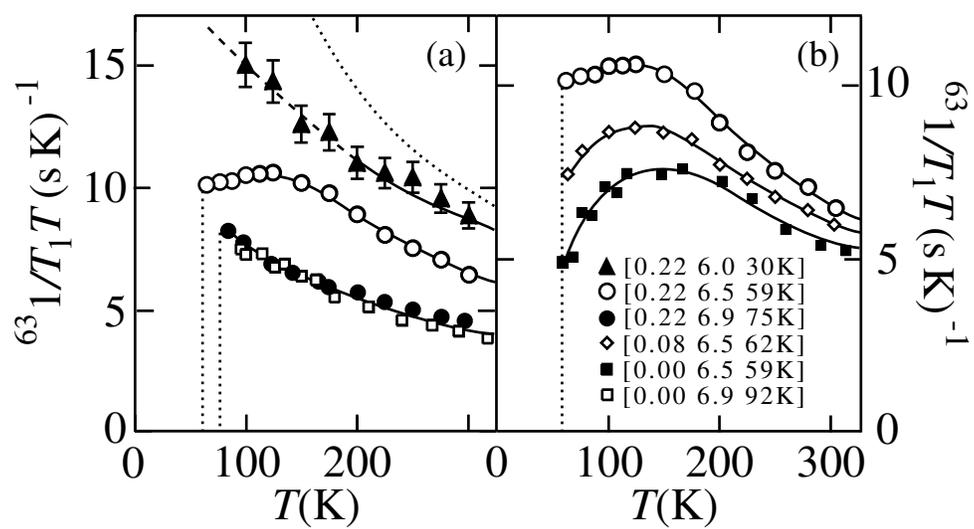
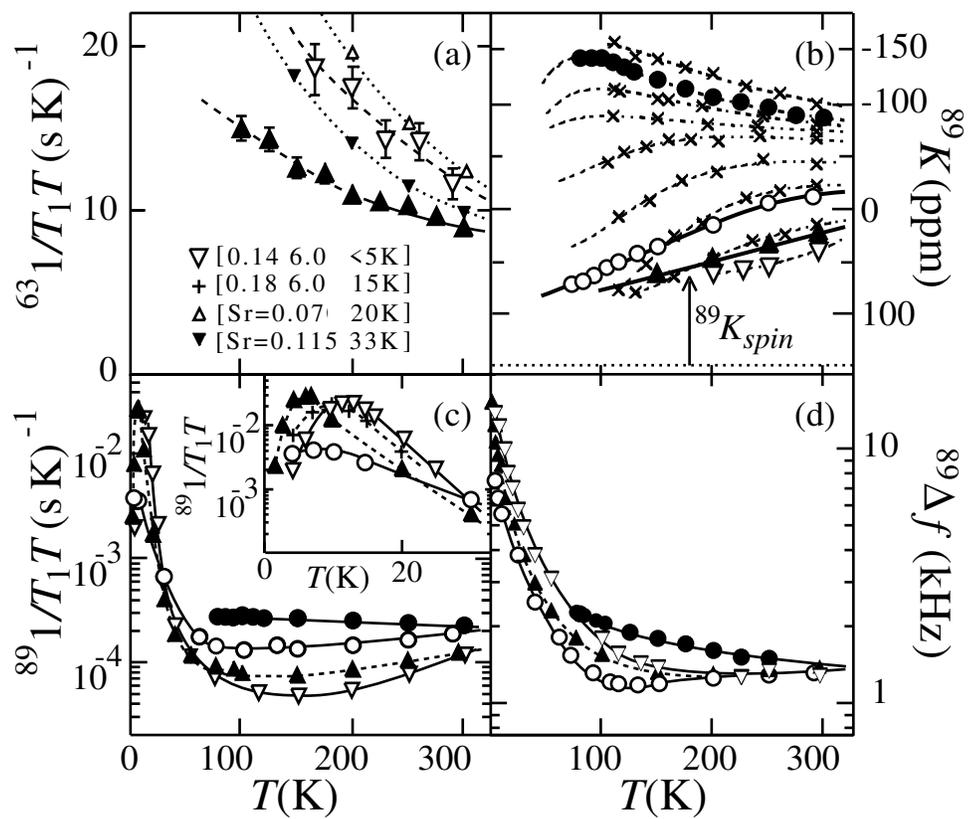
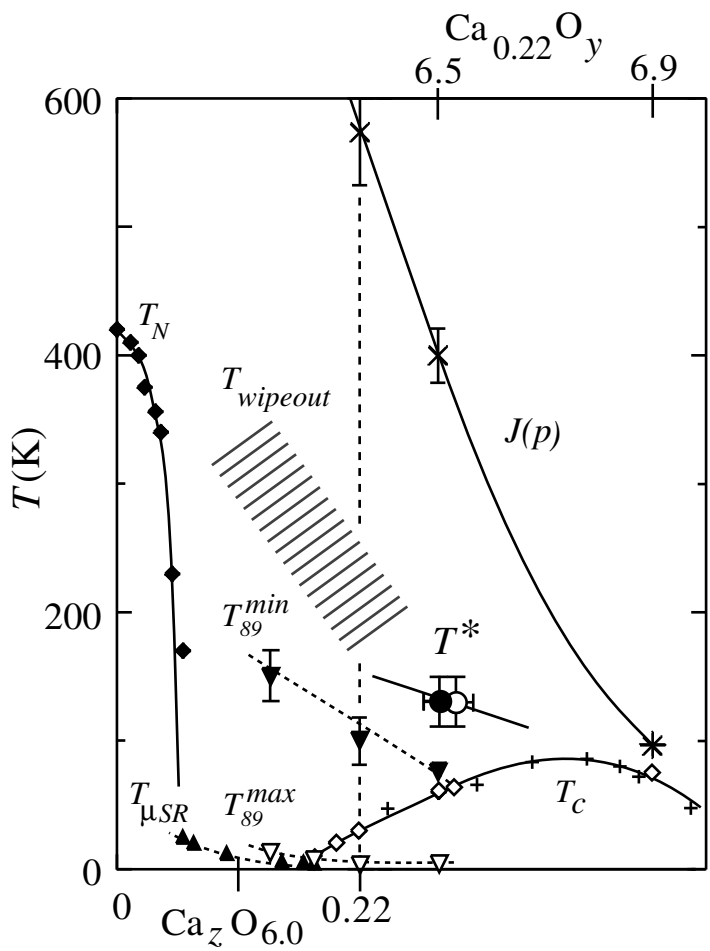
\end{document}